\documentclass{article}

\usepackage{arxiv}

\usepackage[utf8]{inputenc} 
\usepackage[T1]{fontenc}    
\usepackage{url}            
\usepackage{booktabs}       
\usepackage{amsfonts}       
\usepackage{nicefrac}       
\usepackage{microtype}      
\usepackage{lipsum}		

\usepackage{graphicx} 
\usepackage{booktabs} 
\usepackage{caption}
\usepackage{graphicx}
\usepackage{tikz}
\usetikzlibrary{shapes.geometric, arrows}
\usepackage{verbatim}  
\usepackage{subfigure}
\usepackage{color}
\usepackage{algcompatible}
\usepackage{algorithm}
\usepackage[noend]{algpseudocode}
\usepackage[english]{babel}
\usepackage{csvsimple}
\usepackage{booktabs} 
\usepackage{siunitx} 
\usepackage{pgfplotstable} 
\usetikzlibrary{calc}
\usepackage{multirow}
\usepackage{eurosym}
\usepackage{float}
\usepackage{xspace}
\usepackage{pgffor}
\usepackage{xprintlen}
\usepackage{geometry}
\usepackage{url}
\usepackage{tabularx}
\usepackage{amsmath}
\usepackage[export]{adjustbox}
\usepackage{balance}
\usepackage{wrapfig}
\usepackage{epsf}

\usetikzlibrary{shapes.geometric, arrows}

\usepackage[colorlinks,bookmarksopen,bookmarksnumbered,citecolor=red,urlcolor=red]{hyperref}

\usepackage[procnames]{listings} 
\newcommand{\prettysmall}{\fontsize{6.7}{6.7}\selectfont}
\definecolor{gray98}{rgb}{0.98,0.98,0.98}
\definecolor{gray20}{rgb}{0.20,0.20,0.20}
\definecolor{gray25}{rgb}{0.25,0.25,0.25}
\definecolor{gray16}{rgb}{0.161,0.161,0.161}
\definecolor{gray60}{rgb}{0.6,0.6,0.6}
\definecolor{gray30}{rgb}{0.3,0.3,0.3}
\definecolor{bgray}{RGB}{248, 248, 248}
\definecolor{amgreen}{RGB}{77, 175, 74}
\definecolor{amblu}{RGB}{55, 126, 184}
\definecolor{amred}{RGB}{228,26,28}
\usepackage[colorinlistoftodos,prependcaption,textsize=tiny]{todonotes}

\newcommand{\gko}{\textsc{Ginkgo}\xspace}
\newcommand{\oned}{\texttt{regular-1d}\xspace}
\newcommand{\twod}{\texttt{regular-2d}\xspace}
\newcommand{\met}{\texttt{metis}\xspace}
\newcommand{\metis}{\textsc{Metis}\xspace}
\newcommand{\dealii}{\textsc{deal.ii}\xspace}

\title{Evaluating Abstract Asynchronous Schwarz solvers.}

\author{
  Pratik Nayak,\\
  Steinbuch Centre for Computing,\\
  Karlsruhe Institute of Technology,\\
  Karlsruhe, Germany.\\
  \texttt{pratik.nayak@kit.edu} \\
  \AND
  Terry Cojean,\\
  Steinbuch Centre for Computing,\\
  Karlsruhe Institute of Technology,\\
  Karlsruhe, Germany.\\
  \texttt{terry.cojean@kit.edu} \\
  \AND
  Hartwig Anzt,\\
  Steinbuch Centre for Computing,\\
  Karlsruhe Institute of Technology,\\
  Karlsruhe, Germany.\\
  \texttt{hartwig.anzt@kit.edu} \\
}

\date{}

\begin{document}
\maketitle

\begin{abstract}
  With the commencement of the exascale computing era, we realize that the
  majority of the leadership supercomputers are heterogeneous and massively
  parallel even on a single node with multiple co-processors such as GPU’s and
  multiple cores on each node. For example, ORNL’s Summit accumulates six NVIDIA
  Tesla V100’s and 42 core IBM Power9’s on each node. Synchronizing across all
  these compute resources in a single node or even across multiple nodes is
  prohibitively expensive. Hence it is necessary to develop and study asynchronous
  algorithms that circumvent this issue of bulk-synchronous computing for massive
  parallelism. In this study, we examine the asynchronous version of the abstract
  Restricted Additive Schwarz method as a solver where we do not explicitly
  synchronize, but allow for communication of the data between the sub-domains to
  be completely asynchronous thereby removing the bulk synchronous nature of
  the algorithm. 

  We accomplish this by using the onesided RMA functions of the MPI standard. We
  study the benefits of using such an asynchronous solver over its synchronous
  counterpart on both multi-core architectures and on multiple GPU’s. We also
  study the communication patterns and local solvers and their effect on the
  global solver. Finally, we show that this concept can render attractive
  runtime benefits over the synchronous counterparts.
\end{abstract}


\section{Introduction}
\label{sec:intro}

As of today, many of the leadership HPC systems draw a significant portion of
their compute power from accelerators like GPUs. For example, each node of the
Summit supercomputer (Oak Ridge National Laboratory, US) features 6 NVIDIA Volta
V100 accelerators, for a total theoretical peak performance of 40~TFlop/s. In
contrast, the CPU has a theoretical peak performance of 1~TFlop/s.
Accumulating a significant portion of the compute power in highly-parallel
accelerators amplifies the impact of global synchronization points, and there is
an urgent need to move beyond the Bulk-Synchronous Programming Model (BSP). This
is reflected in the US Department of Energy listing the removal of
synchronization bottlenecks as one of the programming challenges for the
exascale era~\cite{amarasinghe2011exascale}.

In consequence, new algorithms should be designed as asynchronous, fault
tolerant, accelerator friendly and agile in terms of adapting to load imbalance.
From the perspective of linear solvers, there are two possible approaches. The
first approach features improving the load balancing of the algorithms and
moving to more asynchronous programming techniques. The second approach is based
on the use of asynchronous iterative methods. While the first approach provides
some performance improvements, it may fail for some algorithms such as the
traditional iterative methods that compute in a lock-step fashion and have
synchronization points boiled into the algorithm. Asynchronous iterative
methods, on the other hand, do not operate in a lock-step fashion, but instead
allow for using the available data without explicitly synchronizing all the
processes at every iteration. This removes the bulk-synchronous character of the
algorithm, and allows to move to a more asynchronous model where each process
trades the global synchronization against the price of possibly computing on old
data. In an abstract view, asynchronous iterative methods trade computation cost
for synchronization cost.

There has been significant research on the characteristics and convergence of
asynchronous methods (\cite{Baudet1978, Frommer2000, Szyld.98} and references
therein), and it has been demonstrated that asynchronous methods can potentially
converge faster than their synchronous counterparts (from a time-to-solution
perspective). Nonetheless, production-ready implementations on HPC architectures
and comprehensive studies of these are a scarce resource, in particular in the
context of machines featuring accelerators such as the Summit HPC system. In
this paper, we implement the abstract Restricted Additive Schwarz algorithm as a
solver for linear systems. For architectures featuring multiple
GPUs in multiple nodes, we show that the asynchronous algorithm is faster than
its synchronous counterpart. Up to our knowledge, this is novel work, and we
hence highlight the following novel contributions:

\begin{itemize}
\item We present an open source implementation of an asynchronous Restricted
  Additive Schwarz algorithm for multi-node systems featuring accelerators.
\item We provide a
  software\footnote[1]{\url{https://github.com/pratikvn/schwarz-lib}} which can
  be used as a testbed for synchronous and asynchronous methods. The software
  can be used on a wide range of hardware architectures, features support for
  both direct and iterative local solves, supports one-sided and two-sided MPI
  communication, and supports both centralized and decentralized convergence
  detection.
\item We show that asynchronous iterative methods can provide significant
  performance improvements over synchronous methods on modern GPU-centric HPC
  systems.
\item We study the effect of different partitioning schemes, overlap size,
  two-sided vs one-sided MPI communication, and centralized vs decentralized
  convergence detection.
\end{itemize}

In Section~\nameref{sec:bgd} we first provide some background on the Schwarz
methods, asynchronous iterative methods, and the
different convergence detection schemes. In Section~\nameref{sec:impl}, we
describe the new algorithm framework that we develop with its capabilities, and
the settings we use in the experimental evaluation that we present in
Section~\nameref{sec:res}. We conclude in Section~\nameref{sec:conc}.

\section{Background}
\label{sec:bgd}

\subsection{Schwarz methods.}
Schwarz methods are a class of domain decomposition methods that were initially
used in \cite{Schwarz1870} as a theoretical tool to show the existence of
solutions for the Laplace equation through the alternating method. Since then,
many variants of the method have found their use as iterative methods, e.g,
~\cite{Dryja1987, Benzi.Frommer.Szyld.01, Ech.Lie.Szy.Tic.18, Dolean2010,
  Blayo2016}, and as preconditioners, e.g, ~\cite{Liu2016,Bonazzoli2017}.

We aim to solve linear systems of the form

\begin{equation}\label{eq:lin_sys}
  Ax=b.
\end{equation}

The general idea of domain decomposition (DD) solvers is to decompose the domain
into distinct subsets so that the local solution for each of those subdomains
can be computed in parallel. To facilitate this and account for the
decomposition, they need to exchange data from each other to reach a global
solution.

\begin{figure}[tb]
  \centering \includegraphics[scale=0.2]{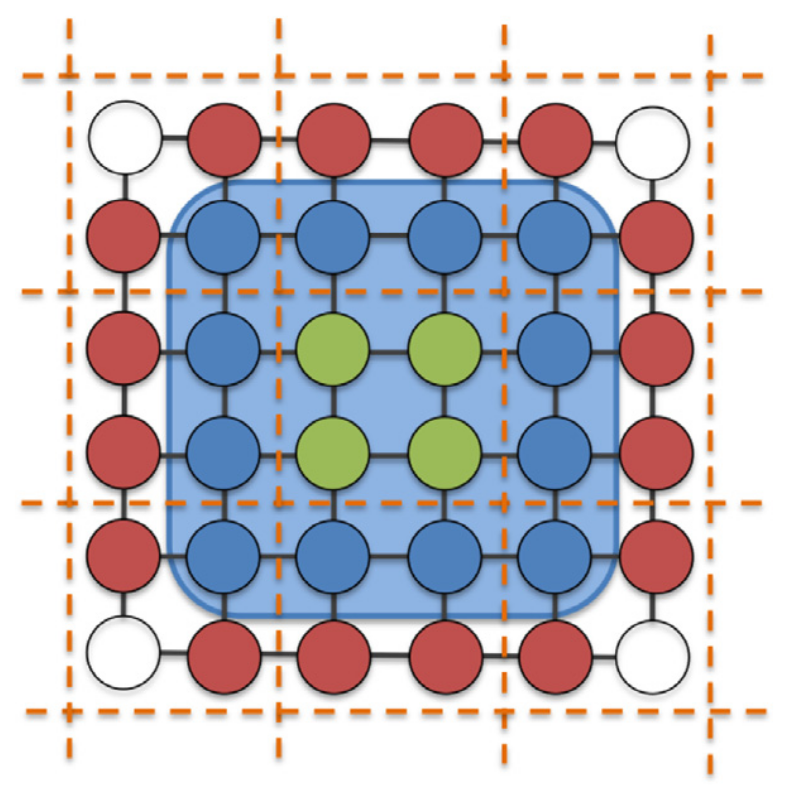}
  \caption{Decomposed 2-D grid with partitioning and overlap. From
    \cite{Yamazaki2019}}
  \label{fig:lap_mesh}
\end{figure}

Consider a two dimensional grid as shown in Figure \ref{fig:lap_mesh}. The
dashed orange lines show the partitioning of the mesh into 3x3 subdomains. The
interior of one subdomain is shown in green. If every point is part of only one
subdomain, the DD method is called a \textit{non-overlapping decomposition}. If, for
example, the subdomain owning the green points includes also the blue points as
variables in the solution of the local problem, then the DD method is said to be
an \textit{overlapping domain decomposition} solver, in this case with overlap $\gamma
=1$. This is the definition of overlap we use throughout this paper. In this
setting, the points marked red are the external interface points through which
the information is exchanged.

In an overlapping additive Schwarz method, the values for the grid points in the
overlap are computed by several subdomains, and to form a unique global solution,
the different locally-computed values need to be weighted~(\cite{frommer1,
  Frommer.Szyld.99}). With a modification proposed by \cite{CaiSarkis99}, the
grid points in the overlap are still considered as variables when computing the
local solution, but discarded afterwards. This modification is called the
\textit{Restricted Additive Schwarz} (RAS) and combines the advantage of converging
faster~\cite{CaiSarkis99} with the advantage of removing any need for weighting
overlap solutions and therewith allowing for collision-free implementation on parallel
computing architectures.

In synchronous Schwarz solvers, the iterations are realized in a lock-step
fashion: the domain is decomposed into subdomains as presented, local solves are
performed in parallel, each subdomain sends the required data to its
``neighbors,'' and all the subdomains wait until data has been received from
their neighbors. Global convergence is achieved when the global solution --
computed after all subdomains solved the local problems -- satisfies a
pre-defined convergence criterion.

Asynchronous Schwarz solvers on the other hand remove the explicit
synchronizations separating the iterations (\cite{Magoules.Szyld.Venet.17,
  MagoulesV18}). Instead, each subdomain proceeds by solving the local problem
using the latest data it received from neighboring subdomains. This has several
critical implications (\cite{Garay.Magoules.Szyld.18.2}): 1) there is no longer
a concept of ``global iterations'' as the distinct subdomains can differ in how
many local solves they completed; 2) local solves can potentially re-compute the
same solution in case no new data was received; 3) the asynchronous Schwarz
convergence and performance is not reproducible as -- theoretically -- each
algorithm execution results in a different subdomain update order and communication
pattern; and 4) the lack of global synchronization points requires alternative
strategies for detecting convergence of the algorithm. Convergence for the
asynchronous Schwarz methods and their variants have been shown, for example in
\cite{Frommer.Szyld.01,Garay.Magoules.Szyld.18.1, Garay.Magoules.Szyld.18.2}.

\subsection{Convergence detection}

Convergence detection in an asynchronous algorithms lacking global
synchronization points is a challenging endeavor. There exists a variety of
strategies to detect convergence of a truly asynchronous algorithm running in a
distributed framework, see, e.g., \cite{Francez1980}, \cite{Mattern1987}, and
\cite{Dijkstra1983} for innovative ideas.

In general, the convergence detection strategies can be classified into two main
categories: centralized and de-centralized convergence detection algorithms. In
centralized termination detection, like the one presented
in~\cite{Yamazaki2019}, a root/leaf paradigm is used, where, each of the leaf
processes reports the convergence status to the root, and the root conversely
sends a termination signal if all of its leaf processes have converged. In the
case of decentralized convergence detection like presented in~\cite{Bahi2005},
there is no such root/leaf paradigm, but each subdomain communicates its
convergence status to its neighbors and a subdomain propagates global convergence
of its solution once it received convergence from all of its neighbors.

\section{Implementation}
\label{sec:impl}

The testbed framework we develop for asynchronous iterative methods is written
in C++ and features a modern C++ design. The framework is intended to be
customizable and extensible, and features multiple options which can be tweaked
or easily added to explore the benefits of various techniques and concepts in
the context of asynchronous iterative methods.

Core of the framework is the algorithm detailed in
Section~\nameref{sec:impl:alg}. This algorithm is generic enough to be reused
for other asynchronous iterative methods since it only calls different
sub-components, which are the \texttt{Initialize} interface detailed in
Section~\nameref{sec:impl:init} and the \texttt{Solve} interface detailed in
Section~\nameref{sec:impl:solve}. The \texttt{Communicate} interface, used in
both the \texttt{Initialize} and the \texttt{Solve} step, supports MPI one-sided
and two-sided communication and is detailed in Section~\nameref{sec:impl:comm}.

The framework we develop interacts with multiple external tools to provide
specific functionalities, among which are \metis~(\cite{metis}) for
partitioning, \dealii~(\cite{dealii}) for generating problems, and the
\gko~\cite{ginkgo} framework for local solvers, generic data management and
certain execution management features as presented in
Section~\nameref{sec:impl:gko}.

\subsection{Core Algorithm}
\label{sec:impl:alg}

Algorithm~\ref{alg:seq} shows the steps involved in the RAS solve. The
initialization step handles the setup and initialization of the solver, while
the solve step handles the local solves, communication, the convergence
detection, and the termination of the solver. The timings and performance data
we report in Section~\nameref{sec:res} account only for the solve step as the
initialization and setup is a constant, non-repeating part of the RAS solver.

\begin{algorithm}[H]
  \begin{algorithmic}[1]
    \Procedure{iterative\_solution}{$A,x,b$}
    \Procedure{initialization\_and\_setup}{} \State Partition matrix \Comment{
      Regular / objective based } \State Distribute data \State Initialize data
    \EndProcedure \Procedure{solve}{} \While{$iter < max\_iter$ or until
      convergence} \State Locally solve the matrix \Comment{Iterative / Direct}
    \State Exchange boundary information \State Update boundary information
    \State Check for Convergence \Comment{ Centralized/Decentralized } \EndWhile
    \State{Gather the final solution vector} \EndProcedure \EndProcedure
  \end{algorithmic}
  \caption{Schwarz Iterative solver} \label{alg:seq}
\end{algorithm}

\subsection{Initialization and setup}
\label{sec:impl:init}

The initialization and setup step consists of three main parts: 1. the
generation and partitioning of the global system matrix; 2. the
distribution/generation of the local subdomain matrices and right hand sides;
and 3. the setup of the communication -- which is detailed in
Section~\nameref{sec:impl:comm} as it is a separate component impacting both the
initialization and the solve characteristics.

\subsubsection{Partitioning}
\begin{figure}
  \centering \subfigure[Metis]{
    \includegraphics[scale=0.13]{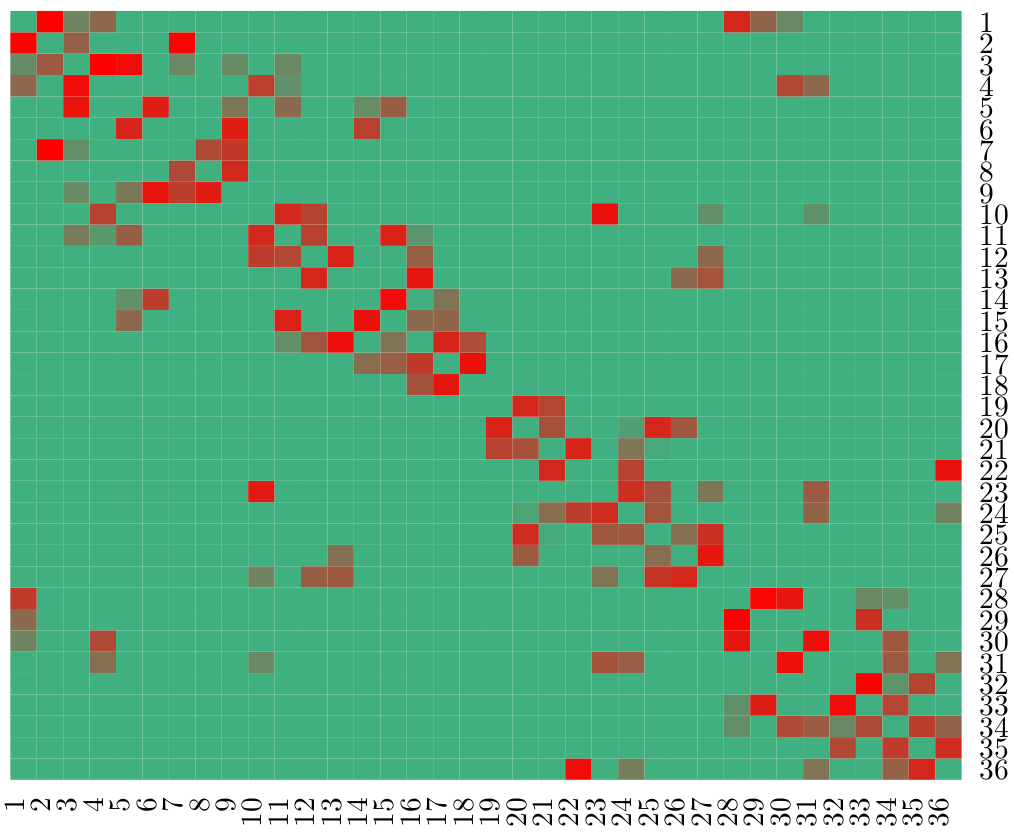} }
  \subfigure[Regular 1D]{
    \includegraphics[scale=0.13]{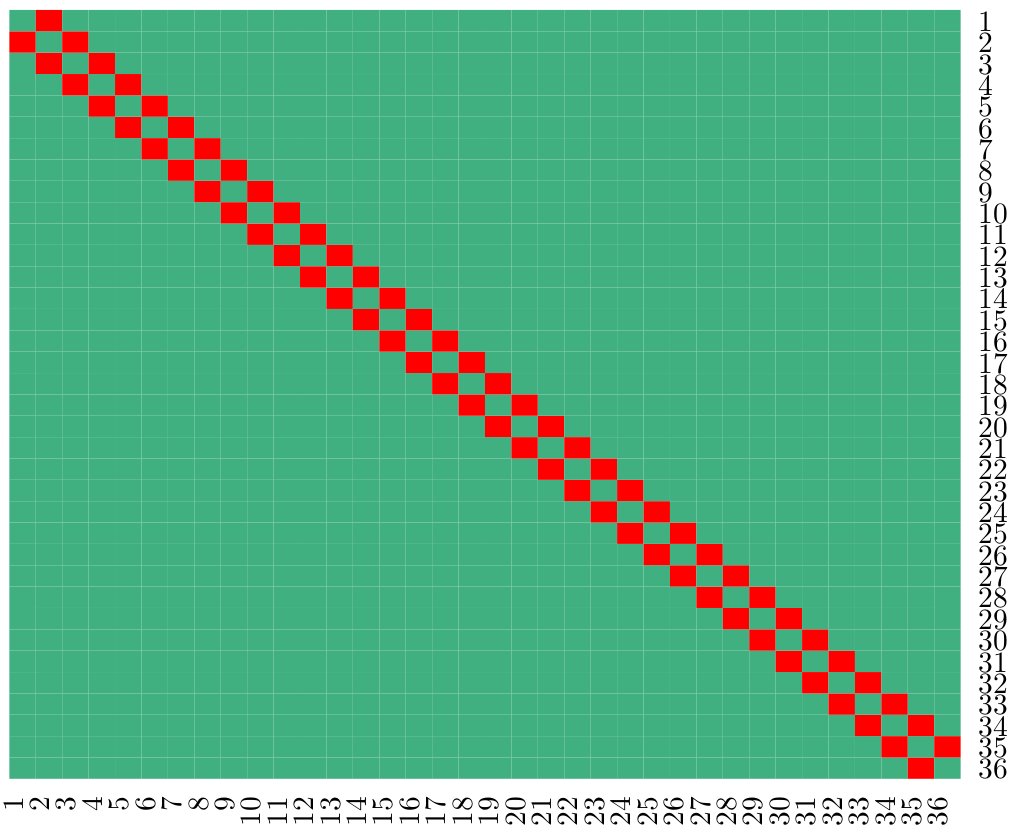} } \subfigure[Regular
  2D]{ \includegraphics[scale=0.13]{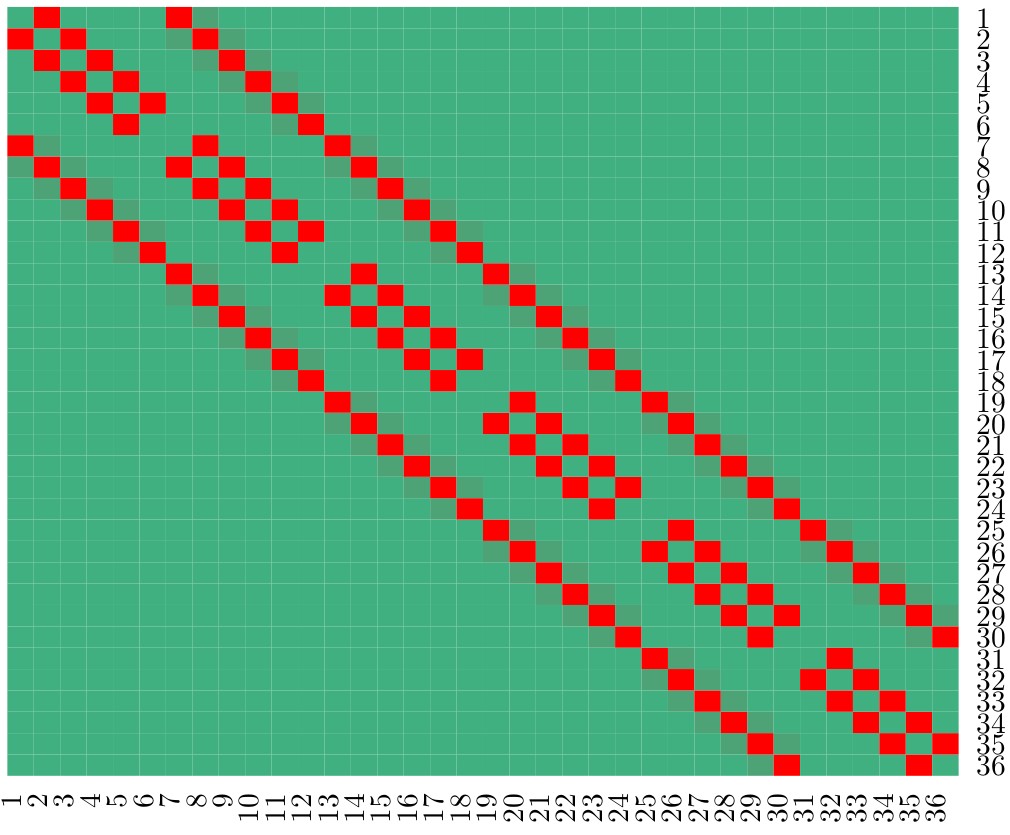} }
  \caption{Communication patterns resulting from different partitioning schemes.
    Subdomain IDs are indicated on the bottom and the right of the communication
    patterns.} \label{plot:part}
\end{figure}

The framework we develop currently supports three partitioning schemes.
Figure~\ref{plot:part} shows the communication patterns that arise for the three
partitioning schemes as heatmaps for a 2D Laplace problem discretized using a 5
point stencil. The communication volume each process receives from its neighbor
is governed by the partitioning scheme. The color-fill in the cells indicate the
amount of communication between the subdomains with the row column pair. Green
indicates no communication and red indicates a significant data volume being exchanged.

In a one-dimensional \texttt{regular1d} partitioning, each subdomain has at most
two neighbors and a very regular communication pattern. At the same time, it is
not very efficient as the information propagation is
slow: For $N$ subdomains, the rate of information propagation, defined here as
the amount of solver iterations needed for data exchange between the two
farthest subdomains, is $N-1$.

In a two-dimensional \texttt{regular2d} partitioning, each subdomain can have at
most 4 neighbors. This allows for faster information propagation at the cost of
increased communication compared to the \texttt{regular1d} scheme.

On the other hand, using the \texttt{metis} partitioning scheme, each subdomain can
have multiple neighbors, and therefore the "farthest distance" can be
significantly smaller.

\subsubsection{Local subdomain matrices}

The partitioning scheme assigns each grid point to a subdomain (or multiple, if
the grid point is part of the overlap). This information is used to assemble the
local subdomain matrix and an interface matrix that is used to communicate
between the different subdomains using a SpMV formulation.

We store all the matrices in the CSR format, which is a popular matrix format
for handling sparse matrices, and which has been shown to perform well for a
given generic matrix. We note that as we use the \gko library, we are not
restricted to this particular format since we can use any other format supported
by \gko (e.g. COO, ELL, Hybrid)~\cite{ginkgo_spmv2019}.

\subsection{Solving the local problems}
\label{sec:impl:solve}

\subsubsection{Local solution}

Each subdomain has its own local matrix and right hand side, and can thus
independently solve the local problem. The local solution can be computed with a
direct solver (computing a LU/Cholesky factorization of the local system once
and solving triangular systems in each iteration), or using an iterative solver
such as CG, GMRES, BiCGSTAB, etc.

In the experimental evaluation, we focus on direct local solves and use the
Cholesky factorization available in the CHOLMOD library (part of the SuiteSparse
package)~\cite{cholmod}. The arising triangular systems are handled with
\gko{}'s triangular solve -- which in turn interfaces to NVIDIA's cuSPARSE
library~\cite{cusparsetrsv}. Using CUDA version 10.1, the
\texttt{cusparse\_csrsm2} sparse triangular solve routine is based on the
level-set strategy and capable of handling multiple right hand sides.

\subsubsection{Convergence detection}
The framework we develop supports two convergence detection strategies: a
centralized (tree-based) convergence detection mechanism and a decentralized
strategy.

\textbf{Centralized, tree based:} This implementation follows the idea presented
in~\cite{Yamazaki2019} in which the children processes detect local convergence
and pass this result to their parent processes. Once the main root has received
such a message, it broadcasts a termination to all its children to terminate the
iterative process.

The local convergence criterion for each of the subdomains is
\begin{equation} \label{eqn:local_conv} ||\tilde{r}_p||^2_2 < \tau^2
  ||\tilde{b}_p||^2_2,
\end{equation}

where $\tilde{r}_p$ is the local residual vector with the values of the subdomain
$p$ (including those in the overlap), $\tau$ is the global solver tolerance, and
$\tilde{b}_p$ is the local right hand side. The global convergence criterion
is $$||r|| < \tau ||b||$$ where $r$ is the global residual vector and $b$ is the
global right hand side. Once all the subdomains have satisfied their local
convergence criteria, we additionally verify that the global convergence
criterion is satisfied post termination.

\textbf{Decentralized, leader based:} This implementation follows the idea
presented in~\cite{Bahi2005}. In this detection algorithm, each subdomain
considers the number of neighbors that have sent a local convergence message.
Once all neighbors have confirmed convergence, the subdomain broadcasts the
global convergence flag to all its neighbors, which in turn broadcast it to
their neighbors and so on. The local convergence criterion is the same as in
\eqref{eqn:local_conv} and, as in the previous algorithm, the processes first
detect their local convergence status and propagate this to their neighbors. 

\subsection{Communication}
\label{sec:impl:comm}
The communication class in our framework handles the communication between the
subdomains. It has two main functionalities. First is the setup and allocation of
all the MPI windows and the communication buffers that handle the
communication between the subdomains. This is shared with the
\texttt{Initialization} class. Second is the detection of convergence and
termination of the solver. This is shared with the \texttt{Solver} class. We now
explain the setup for the two cases:

{\bf Synchronous Schwarz setup:} With the synchronous
version, each step in the solve procedure in Algorithm \ref{alg:seq} is
performed in a lock-step fashion. We associate each subdomain to one MPI rank. A
subdomain here refers to the computational unit that performs a local solve and
communicates the required data to its ``neighbors''. Currently, the subdomains
in one RAS solve are restricted to either only GPUs or only CPUs.

We use MPI point to point communication to communicate between the different
subdomains. We use the CUDA-aware MPI to directly transfer the buffers between
GPUs of different subdomains instead of intermediately staging them on the CPU.
This allows for faster communication as the latencies from the CPU side are
hidden (particularly for GPUs attached to the same socket and communicating via
the CUDA NVLINK technology~\cite{nvlink}). We use the non-blocking point to
point functions \texttt{MPI\_Isend} and \texttt{MPI\_Irecv} to send and receive
data between the subdomains. Before sending, the data is first packed and after
receiving the target subdomain unpacks it. An \texttt{MPI\_Wait} is called by
each subdomain to wait for the \texttt{MPI\_Isend} from its neighboring
processes. This results in explicit synchronization points separating the
Schwarz iterations.

{\bf Asynchronous Schwarz setup:} In the asynchronous version, each step in the
solve procedure in Algorithm \ref{alg:seq} is executed without synchronizing
with neighbors. The MPI-2 standard~(\cite{mpi2standard}) introduced the Remote
Memory Access (RMA) functions that allow remote MPI ranks to access dedicated
buffers without explicit synchronization. Each process can ``put'' its data into
the buffer of its neighbors through a ``window'' using the RMA function
\texttt{MPI\_Put}. To make sure that the operations are completed, a ``flush''
is called on the windows. Avoiding global synchronization points makes this
algorithm truly asynchronous in terms of the communication.


\subsection{The \gko framework}
\label{sec:impl:gko}

The \gko numerical linear algebra library is a node-level high performance
sparse linear algebra library featuring high performance kernels for different
back-ends such as multicore (OpenMP), AMD GPUs (HIP) and Nvidia GPUs (CUDA).
\gko provides the user with simple and powerful linear operator abstraction to
interface various solvers, preconditioners, and matrix operations. In the
experimental analysis, we use \gko objects for all arrays, matrices, and
vectors. This enables us to leverage \gko{}'s executor concept -- it allows to
easily move from one hardware architecture to another without code changes
except for choosing another execution space. As we are focusing on the
architecture of the Summit supercomputer, we can run our Schwarz framework on
the V100 GPUs using \gko{}'s CUDA executor.

\section{Experimental Assessment}
\label{sec:res}




\begin{figure}[tb]
\centering
\includegraphics[width=.75\columnwidth]{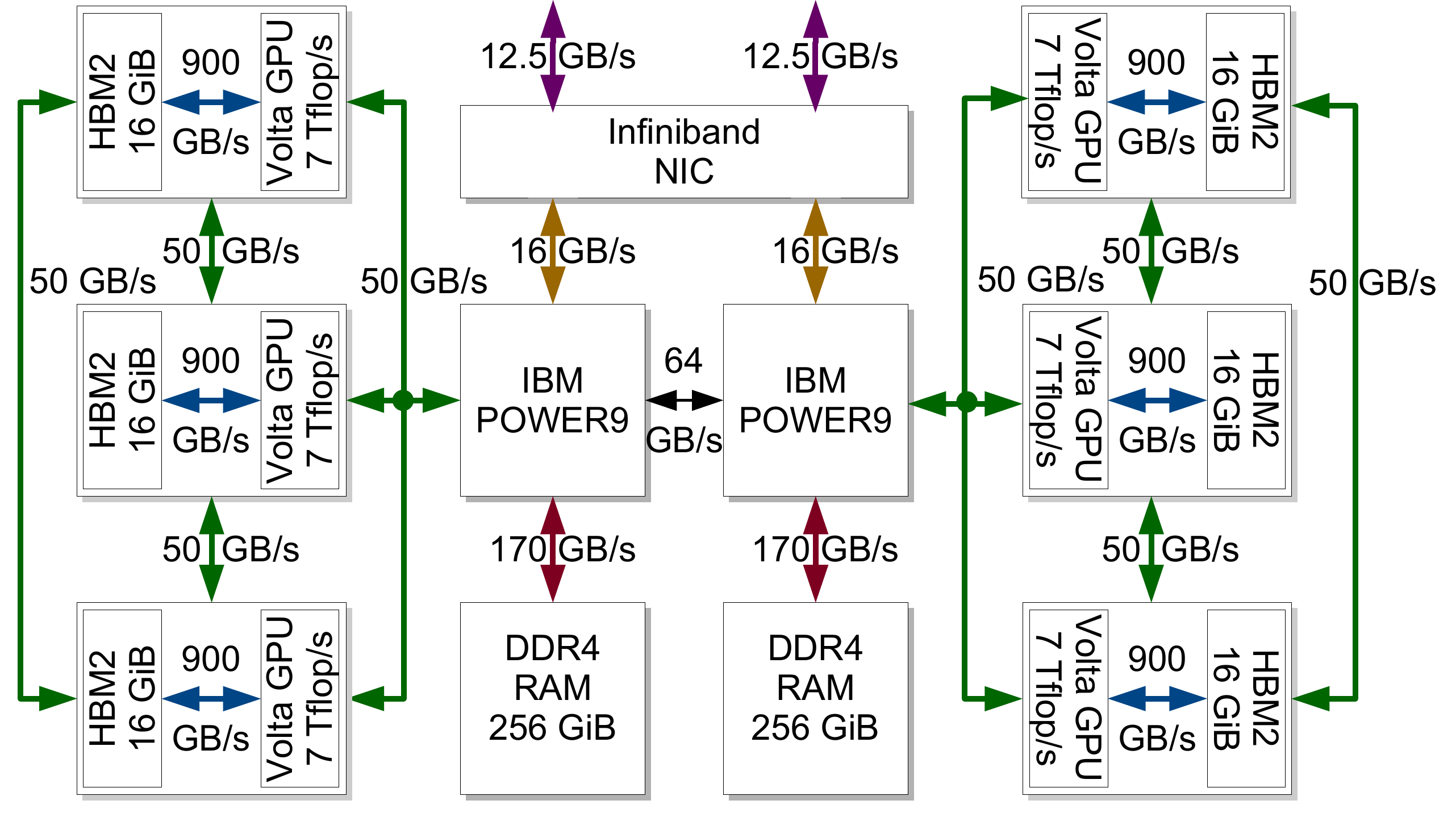}
\caption{On-node hardware connectivity diagram for the Summit supercomputer at
  Oak Ridge National Laboratory based on IBM POWER9 processor and NVIDIA Volta
  GPU accelerators.}
\label{fig:summit}
\end{figure}

\subsection{Hardware and software setup}

For the experimental analysis we focus on the 2D Laplace problem discretized
using a 5 point stencil, $\mathcal{S} = \{-1,-1,4,-1,-1\}$ with offsets
$\{-N,-1,0,1,N\}$ and $N$ being the number of grid points in each discretization
direction. The resulting linear system is very regular, therewith promoting good
load balance and making it particularly difficult for asynchronous methods to
outperform synchronized counterparts.We use a random RHS, and define the
convergence of the RAS solver once the relative residual norm drops below
$1e-7$. We would also like to mention that the RAS solver is capable of solving
a generic linear system when coupled with a generic local solver such as
LU/GMRES.

The benchmark runs are conducted on the Summit supercomputer at the Oak Ridge
National Lab. Each node composes of two IBM Power 9 CPUs and 6 NVIDIA Tesla V100
GPUs. The 3 GPUs connected to the same socket are directly connected to each
other via a NVLINK brick with a 50GB/s bi-directional bandwidth, see
Figure~\ref{fig:summit}~\cite{summit}. Each V100 features 16GB of High Bandwidth
Memory (HBM2). In total, the Summit supercomputer contains more than 4,600 nodes
of this kind. The nodes are connected via a dual-rail EDR InfiniBand network
(non-blocking fat tree) providing a bandwidth of 23GB/s. Hence, in the scaling
analysis, we need to consider the different bandwidth capabilities. The fastest
communication between two GPU's is through the direct NVLINK (50GB/s) that
connects GPU's of the same node. The second fastest communication is the
off-socket on-node , but still on node: NVLINK (50GB/s) -- Inter-socket (64GB/s)
-- NVLINK (50GB/s) path. The slowest communication is between GPU attached to
different nodes, which requires going through the ``slow'' Interconnect
(23GB/s).

This paper's codebase uses \gko as a central building block for the RAS solver.
Overall, local on-node (on-GPU) operations use \gko functionality, while the
communication is handled via MPI. As we perform our experiments on the Summit
system, we use the IBM Spectrum MPI library (a flavor of OpenMPI). Related
work~\cite{Yamazaki2019} analyzes the impact of using different MPI
implementations including the performance of the one-sided RMA functions.

Investigating asynchronous methods like the asynchronous RAS solver is
challenging for several reasons. First, due to the non-deterministic behavior,
we are generally unable to predict or reproduce results and effects. At best, we
can report statistical data and draw weak assumptions. In consequence, all data
we present for the asynchronous RAS in this paper is averaged over 10 runs.
Second, in asynchronous iterative methods, the concept of iterations does no
longer exist. Removing explicit synchronization points it is possible that some
solver parts (i.e. subdomains) have already completed a high number of local
solves while other parts did not yet complete a single local solve. To reflect
this challenge, we base all performance data we report on the asynchronous RAS
on the time-to-solution. This is the total time until global convergence is
detected -- which brings us to a third challenge: without global synchronization
points, it is difficult to detect global convergence. The framework we develop
supports both centralized and decentralized convergence detection, and while we
analyze the performance effect of these strategies at the end of this section,
we use decentralized convergence as default setting.

\begin{figure}[h]
  \centering
  \begin{minipage}[b]{0.45\textwidth}
    \subfigure[Runtime breakdown for the averaged subdomain update]{
      \includegraphics[scale=0.16]{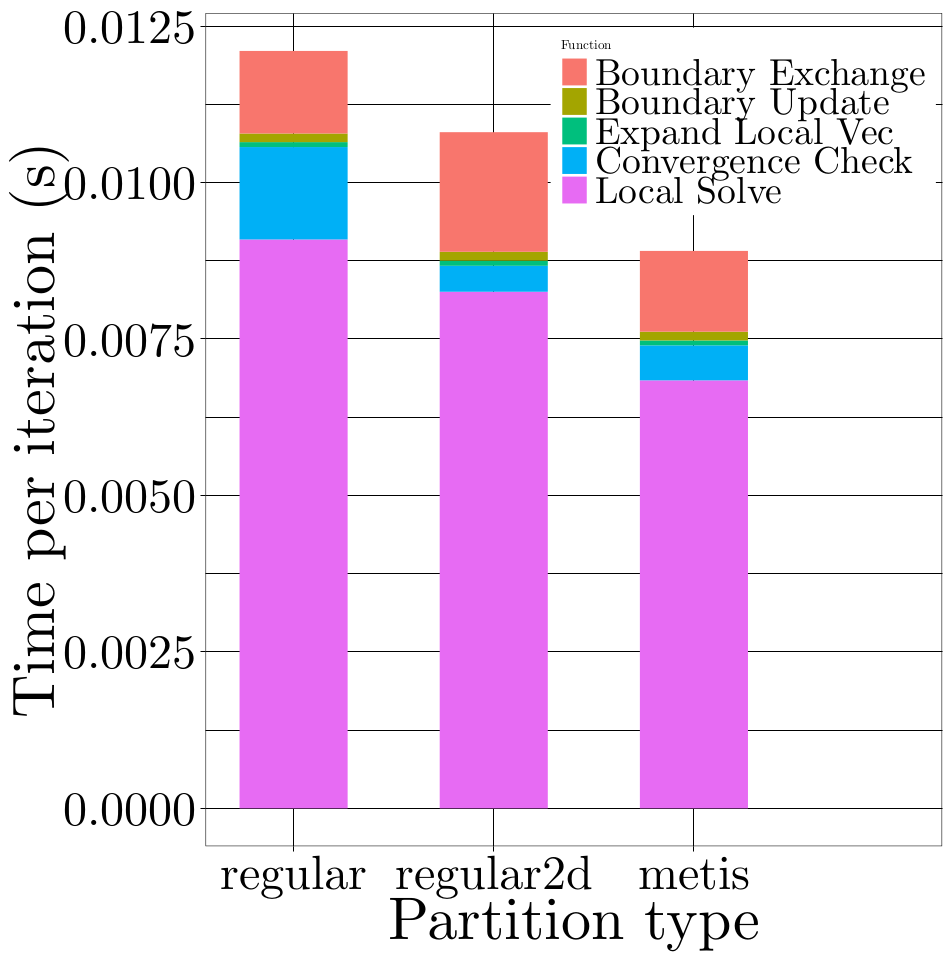}
      \label{fig:part:func}
    }
  \end{minipage}
  \begin{minipage}[b]{0.45\textwidth}
    \subfigure[Speedup - Asynchronous vs Synchronous]{
      \includegraphics[scale=0.21]{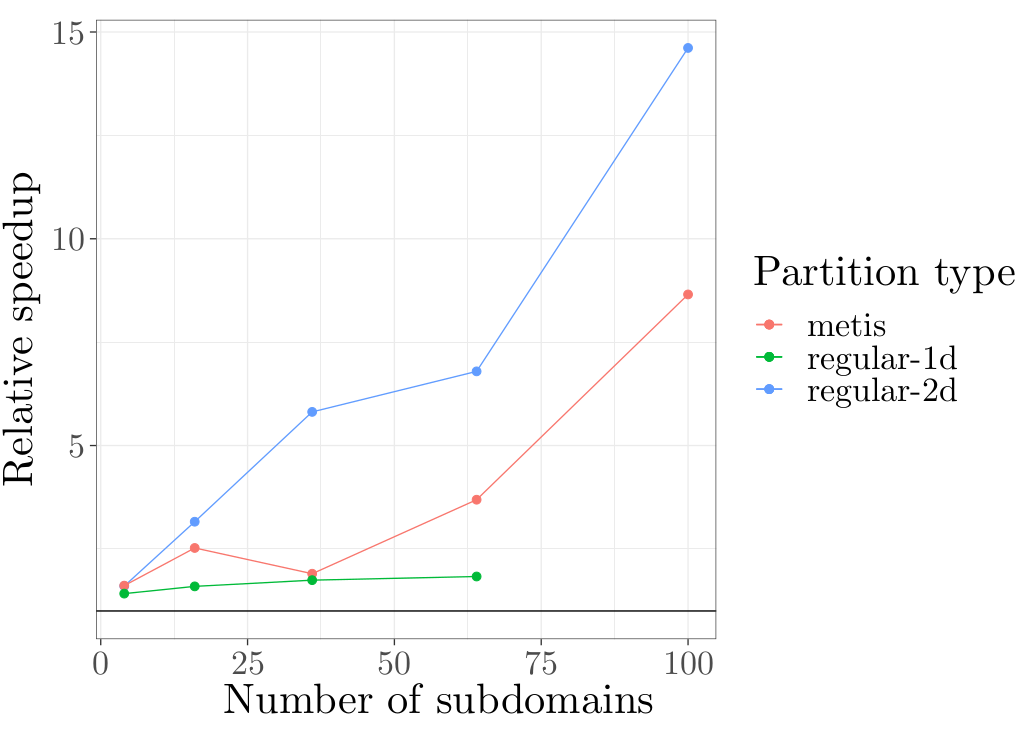}
      \label{fig:part:speedup}
    }
  \end{minipage}
  \caption{Comparing the different partitioning strategies. (a) shows the time
    spent in the different functions (sync),
    and (b) shows the speedup of the asynchronous version over the synchronous
    version.}\label{fig:part}
\end{figure}

\subsection{Comparing partitioning strategies}

Figure~\ref{fig:part:func} visualizes the runtime needed for one synchronous RAS
iteration using two-sided MPI communication averaged over all the subdomains
(problem with 262144 unknowns divided into 4 subdomains). Different partitioning
strategies are considered. Partitioning affects the solution in two ways: On the
one hand, it affects the fill-in governed by the factorization of the local
subdomain matrices. \twod and \met are better at minimizing the fill-in of the
resulting local matrices when distributing the rows to the subdomains. Hence, we
expect that the local solve is cheaper when using \twod and \met. On the other
hand, it dictates the communication volume between the subdomains. \met is
better in minimizing the edge cut of the adjacency matrix, and the communication
time is reduced compared to \twod.

For a weak scaling analysis, we now fix the local problem size to {4096}
elements per subdomain, and assess the time-to-solution speedup of the
synchronous RAS solver using \twod and \met over \oned for increasing subdomain
counts.

In Figure~\ref{fig:part:speedup} we investigate the effect of abandoning the
synchronization points in the RAS solver. Again, the speedup is based on the
time-to-solution and the asynchronous versions of the partitioning methods are
compared to their synchronous counterparts. The number of elements per subdomain
is set to 4096 and we increase the number of subdomains from 4 to 100. The \oned
partitioning does not scale for a large number of processes. We observe that as
we increase the number of subdomains, the asynchronous version outperforms the
synchronous version for all partitioning methods even reaching a maximum speedup
of about 15$\times$ for a large number of subdomains for the \twod partitioner.
The performance of the \twod partitioner is on average 1.3 $\times$ slower than
when \met partitioner is used for the synchronous version. We would like
to compare the performance of our asynchronous solver with the best parameters
for the synchronous version, and given the fact that \met is a generic
partitioner (applicable to any generic matrix), we choose to run all our
experiments with the \met partitioner.

\begin{figure}[h]
  \centering
  \begin{minipage}[b]{0.45\textwidth}
    \subfigure[Runtime breakdown for the averaged subdomain update \& subdomain
    update count]{ \includegraphics[scale=0.18]{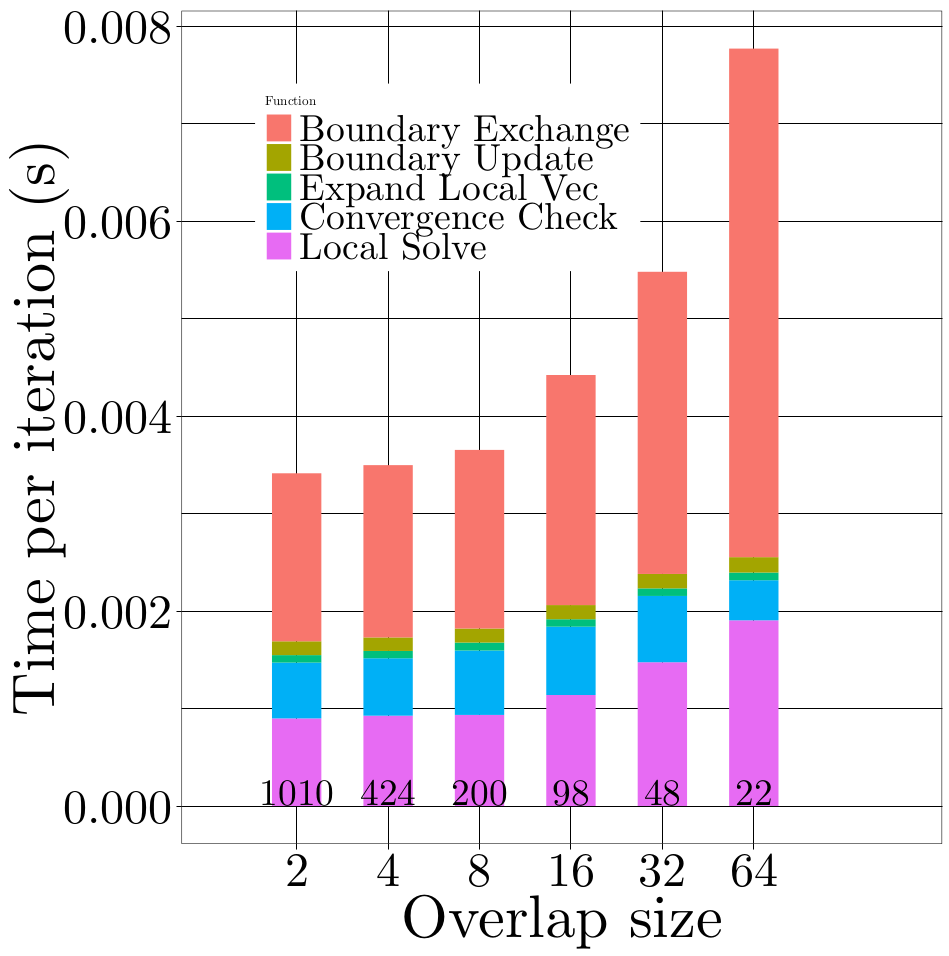}
      \label{fig:overlap:func}
    }
  \end{minipage}
  \begin{minipage}[b]{0.45\textwidth}
    \subfigure[Speedup - Asynchronous vs Synchronous]{
      \includegraphics[scale=0.23]{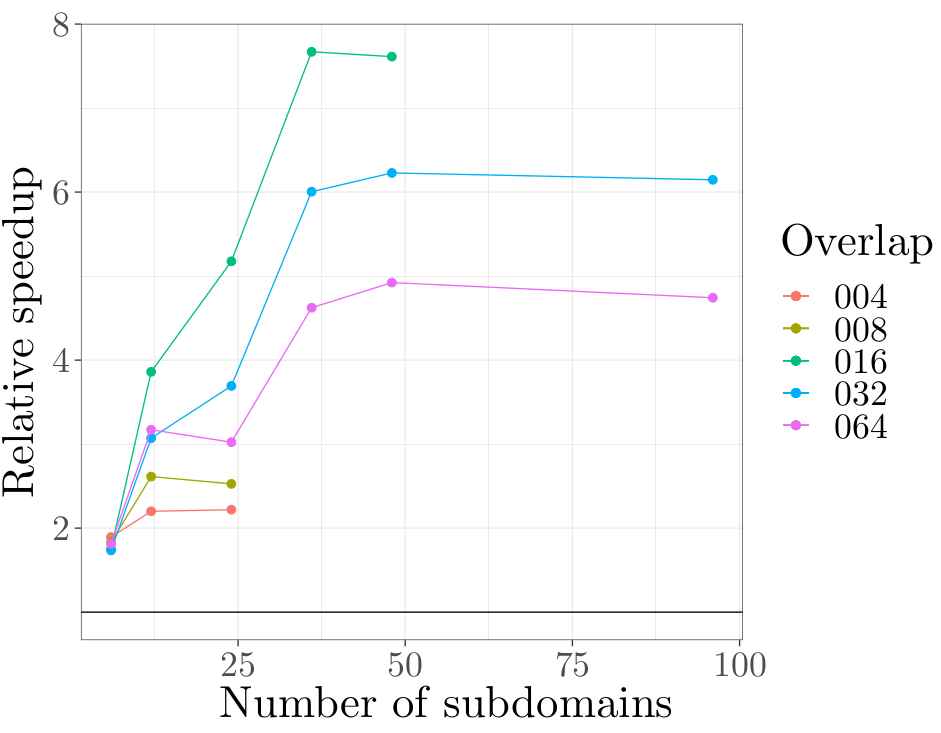}
      \label{fig:overlap:speedup}
    }
  \end{minipage}
  \caption{Comparing the different overlaps between the subdomains. (a) shows
    the time spent in different functions (synchronous version), (b) shows the
    speedup of the asynchronous version over the synchronous
    version.}\label{fig:overlap}
\end{figure}

\subsection{Comparing different overlaps}

Figure~\ref{fig:overlap:func} shows for different overlap sizes the time spent
in the different components of a single (synchronized) RAS iteration and the
total number of RAS iterations needed for convergence. The global problem of
size 16384 is decomposed into 6 subdomains, and only the
overlap is varied. All timings are averaged over the 6 subdomains. A minimal
overlap of 2 elements is required for the RAS method to converge, but a minimal
overlap is inefficient as the information propagation is slow, which results in
the RAS solver needing over 1000 iterations to converge. The local problem size
increases with the overlap, increasing the cost of the local solves. The time
for "Boundary Exchange" also increases with the overlap as more data has to be
communicated. At the same time, the larger overlaps enabling the faster
information propagation promotes much faster convergence, i.e. reduce the number
of needed RAS iterations.
The optimal overlap has to balance between these aspects and depends on the
problem characteristics, the partitioning scheme, and the ratio between compute
power and memory bandwidth of the used architecture. The asynchronous solver
benefits from having a high overlap because higher overlaps reduce the message
cost and do not have the synchronizations that adversely affect the synchronous
version. This is also seen from Figure~\ref{fig:overlap:speedup}, where we
compare the speedup for same overlap size of the asynchronous version against
the synchronous version for varying problem size. The time to converge for the
lower overlap values, namely ``4'' and ``8'' for especially the synchronous
version is high and excluded from this graph. The asynchronous version can
perform much better than the synchronous version when using an overlap of size
``16'' as most of the communication time can be hidden.

\begin{figure}[h]
  \centering
  \begin{minipage}[b]{0.45\textwidth}
    \subfigure[Runtime breakdown for the averaged subdomain update
    (Asynchronous)]{ \includegraphics[scale=0.18]{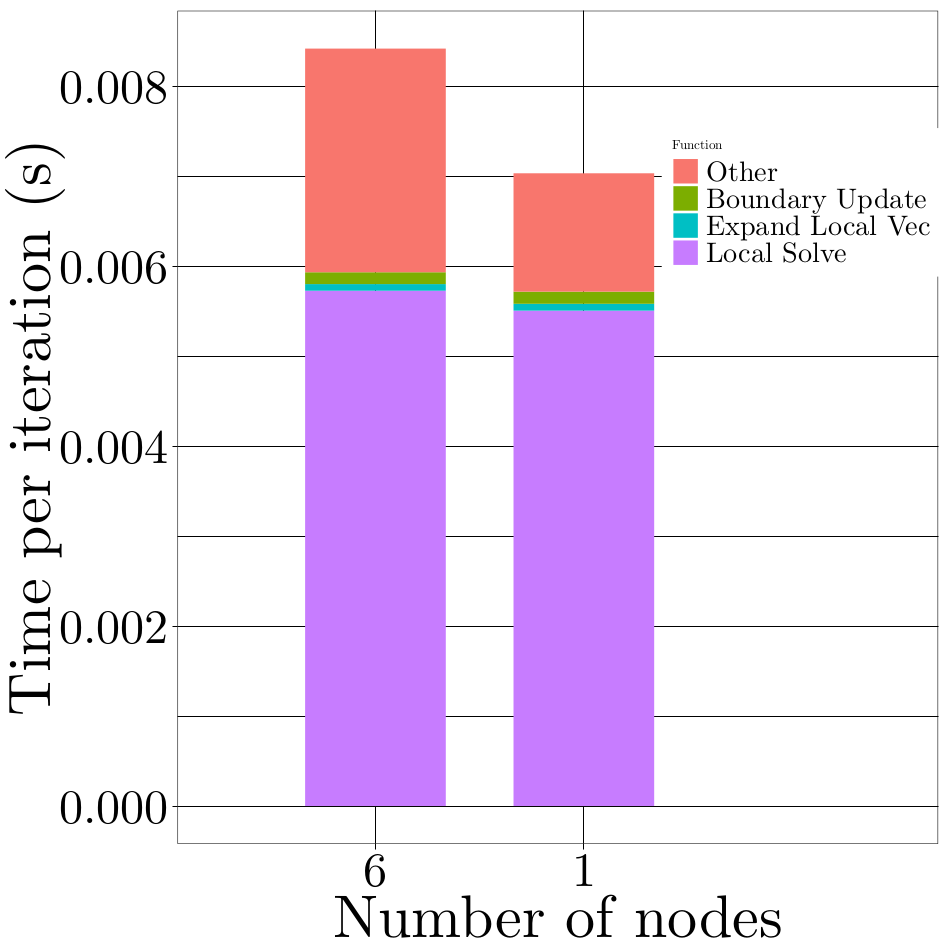}
      \label{fig:node:func}
    }
  \end{minipage}
  \begin{minipage}[b]{0.45\textwidth}
    \subfigure[Scaling Efficiency]{
      \includegraphics[scale=0.22]{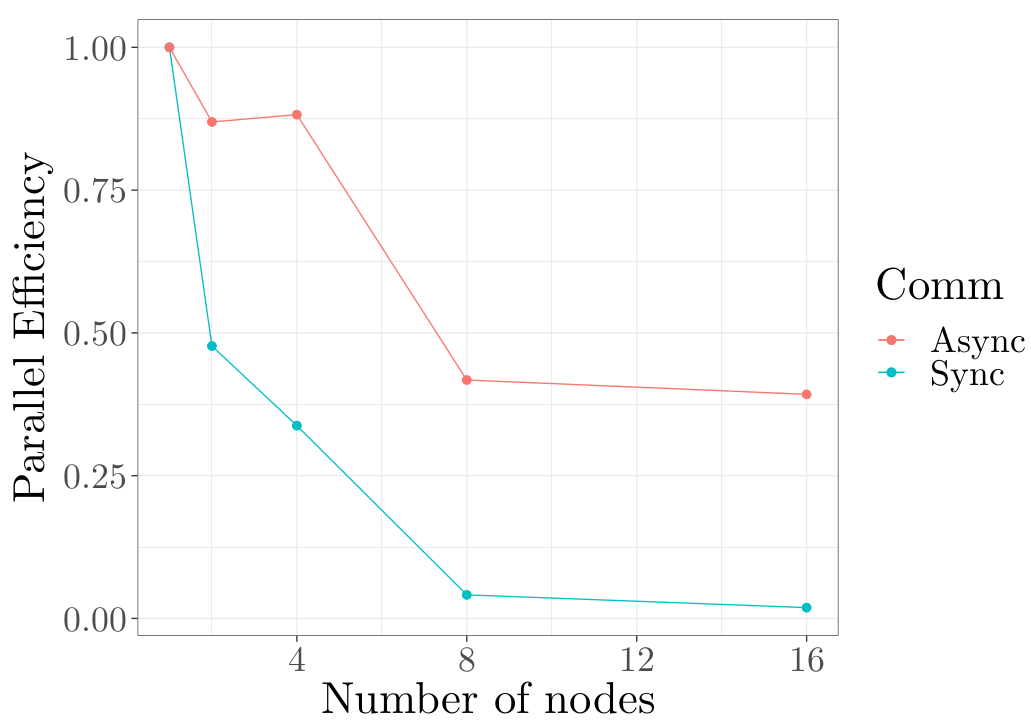}
      \label{fig:node:scale}
    }
  \end{minipage}
  \caption{Comparing scaling on increasing number of nodes. (a) shows the time
    spent in different functions, (b) shows the parallel scaling efficiency
  }\label{fig:node}
\end{figure}

\subsection{Scalability study}

Before investigating the scalability of the synchronous and asynchronous RAS
solver, we study the effect of the internode vs NVLINK bandwidths. In the
asynchronous RAS, we can not isolate the communication operations, but we can
measure the overall time-to-solution and the time each subdomain spends on
solving the local problem. Figure~\ref{fig:node:func} visualizes the time spent
in different routines when using 6 subdomains, either executed in 6 GPUs
attached to the same node, or in 6 GPUs attached to distinct nodes. The averaged
run times over all the subdomains reveal that the local solve cost remains
unaffected by the hardware configuration, but the overall time increases by
about 20\% when communicating over the InfiniBand fabric.

Figure~\ref{fig:node:scale} presents a weak scaling analysis for the
asynchronous RAS using the same problem size per subdomain (4096 elements per
subdomain) and increasing the node count (always 6 subdomains per node, each
subdomain on one GPU). We note that increasing the number of subdomains comes
with two effects: 1) The computational cost increases (so the communication);
and 2) the different subdomain count and partitioning typically results in a
different iteration count for the RAS method. Unfortunately, for asynchronous
methods, these effects are difficult to isolate, and in this study we focus on
the overall time-to-solution. The speedup therefore combines an algorithmic
speedup (faster/slower convergence) with an implementation/architecture speedup
(use of parallel resources). We observe that due to the synchronizations, the
synchronous RAS does not scale well. The asynchronous RAS on the other hand, due
to its asynchronous nature, can overcome these bottlenecks and as the number of
subdomains increases, the parallel efficiency is about 37\% for 16 nodes (96
subdomains on 96 GPUs).

\begin{figure}[h]
  \centering
  \begin{minipage}[b]{0.45\textwidth}
    \subfigure[Runtime Breakdown for the subdomain update]{
      \includegraphics[scale=0.2]{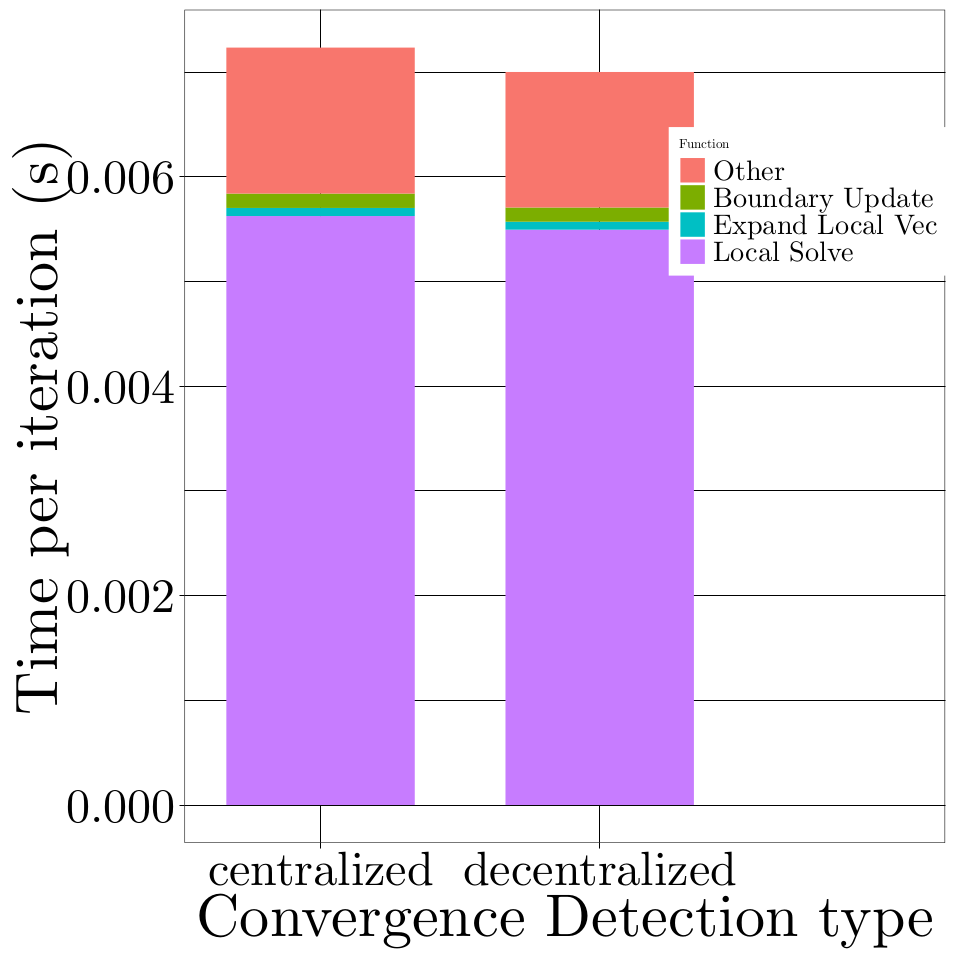}
      \label{fig:conv:func}
    }
  \end{minipage}
  \begin{minipage}[b]{0.45\textwidth}
    \subfigure[Speedup - Decentralized vs Centralized]{
      \includegraphics[scale=0.22]{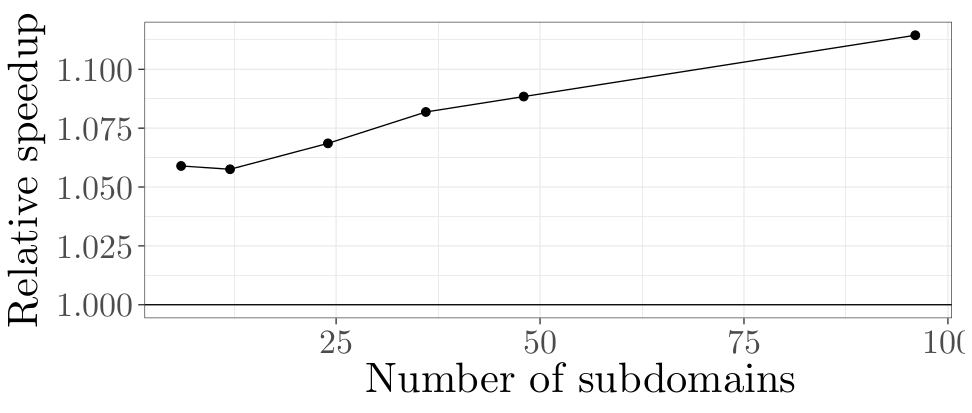}
      \label{fig:conv:speedup}
    } \subfigure[Subdomain update spread]{
      \includegraphics[scale=0.19]{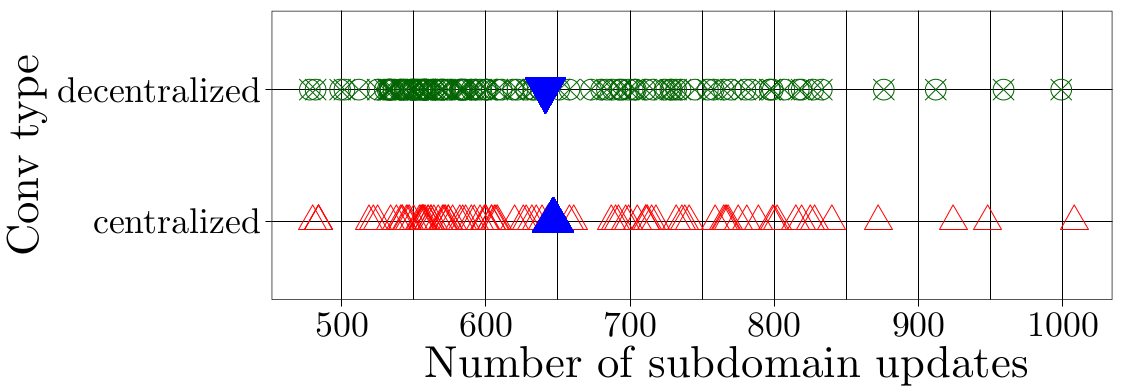}
      \label{fig:conv:numiter}
    }
  \end{minipage}
  \caption{Comparing the centralized and decentralized convergence detection
    algorithms. (a) shows the time spent in different functions, (b) shows the
    speedup of using a decentralized vs using a centralized algorithm and (c)
    shows the number of subdomain updates for the two methods for all the
    subdomains.}\label{fig:convdet}
\end{figure}

\subsection{Comparing convergence detection algorithms}

In Figure~\ref{fig:conv:func}, we compare the centralized and decentralized
convergence detection strategies supported by the developed framework with all
other parameters constant (asynchronous, i.e. one-sided version, \met
partitioner, overlap of 16 elements, problem size 262144). As the number of
subdomains increases, a centralized tree detection has a higher overhead because
everything is centralized through the root, see Figure~\ref{fig:conv:speedup}. A
better understanding of the two convergence detection algorithms can be obtained
from Figure~\ref{fig:conv:numiter} which shows the variance in the number of
subdomain updates (local solves) with the average number of local solves counts
indicated with blue triangles (for 96 subdomains). The decentralized algorithm
is able to detect convergence slightly faster, which also reduces the
time-to-solution compared to the centralized counterpart.

\begin{figure}[h]
  \centering
  \begin{minipage}[b]{0.45\textwidth}
    \subfigure[Runtime breakdown for the averaged subdomain update]{
      \includegraphics[scale=0.2]{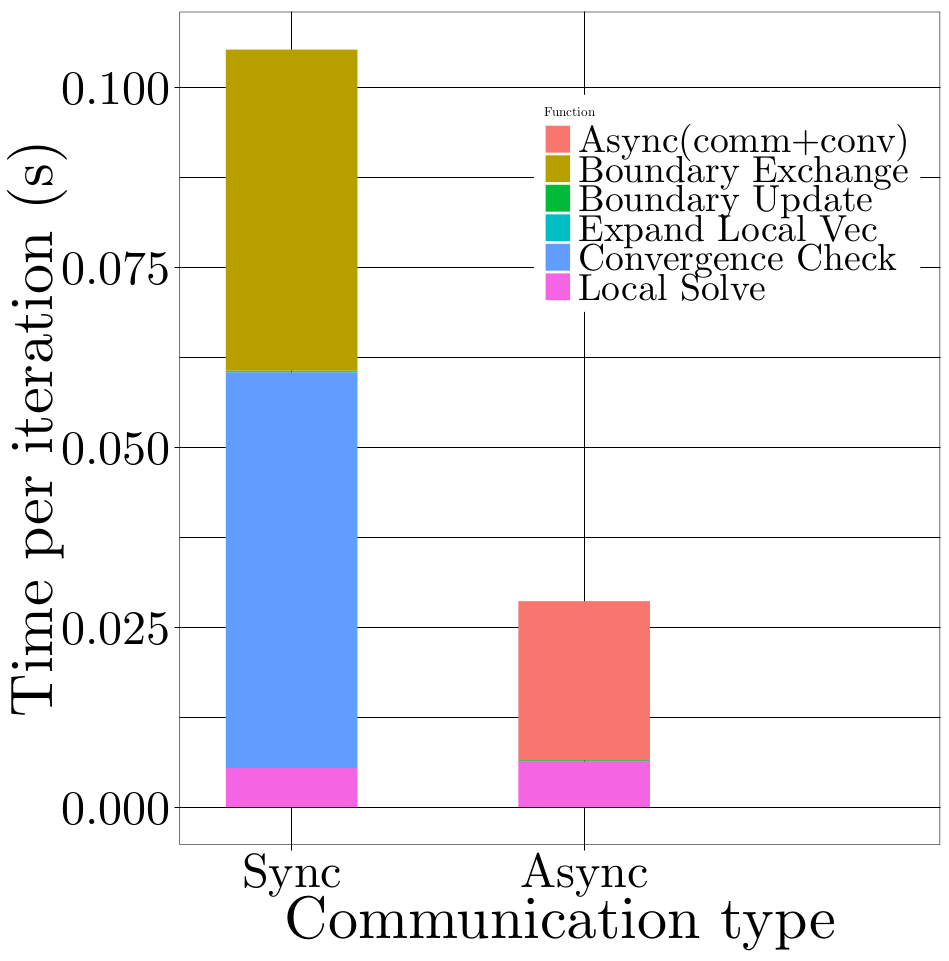}
      \label{fig:comm:func}
    }
  \end{minipage}
  \begin{minipage}[b]{0.45\textwidth}
    \subfigure[Speedup - Asynchronous vs Synchronous]{
      \includegraphics[scale=0.22]{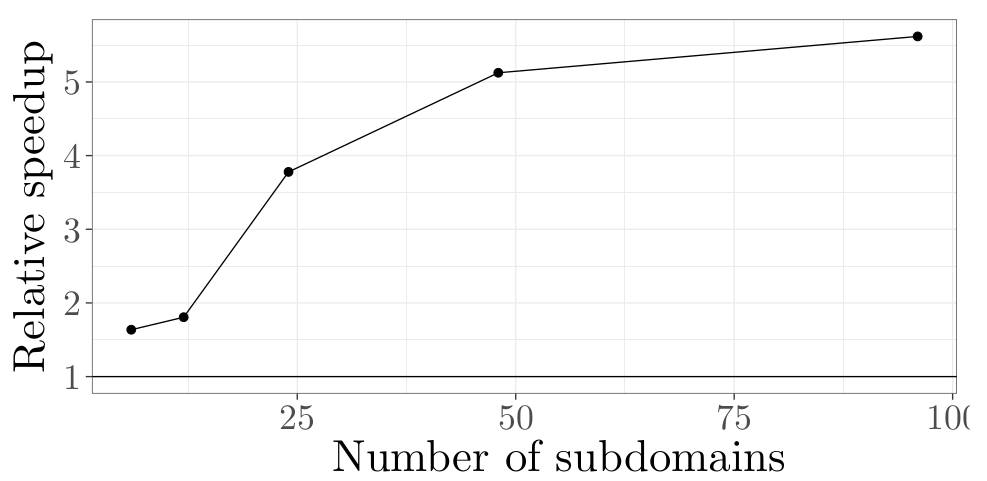}
  		\label{fig:comm:speedup}
  	} \subfigure[Subdomain update spread]{
      \includegraphics[scale=0.19]{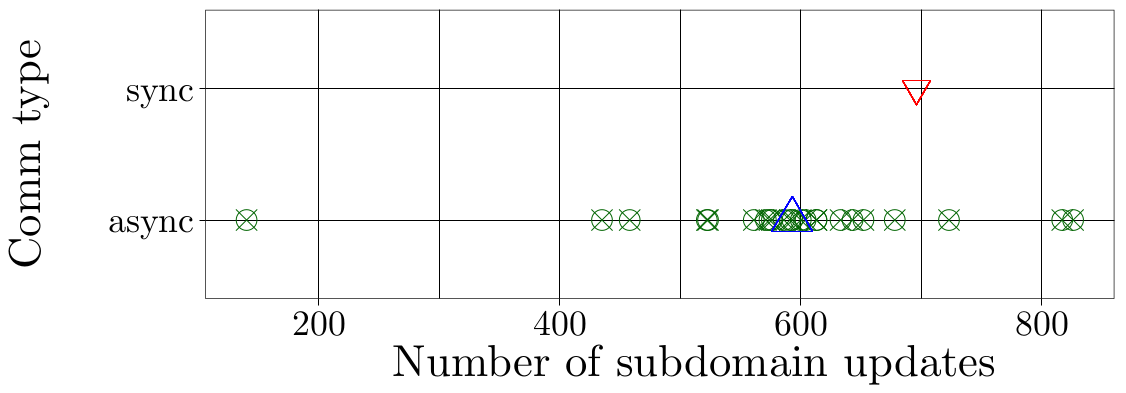}
      \label{fig:comm:numiter}
    }
  \end{minipage}
  \caption{Comparing scaling on increasing number of comms. (a) shows the time
    spent in different functions, (b) shows the speedup of the asynchronous
    version (one-sided MPI) over the synchronous version (two-sided MPI) and (c)
    shows the subdomain update spread for the two versions.}\label{fig:comm}
\end{figure}

\subsection{Comparing two-sided and one-sided communications}

We finally compare the asynchronous RAS solver (one-sided MPI) against the
synchronous RAS solver (two-sided MPI) using the respectively optimal
settings. Figure~\ref{fig:comm:func} compares the runtime breakdown of one
subdomain update of the synchronous RAS solver with the runtime breakdown of
the asynchronous RAS solver (problem size 1M, 36 subdomains, 16 overlap).
Figure~\ref{fig:comm:speedup} relates the time-to-solution speedup of the
asynchronous version over the synchronous version to the subdomain count (every
subdomain handles a problem of size 4096). The speedup increases with the subdomain
count, reaching a speedup of about 5.5$\times$ for 96 subdomains. In
Figure~\ref{fig:comm:numiter}, we visualize for the 36 subdomain case the number
of local solves the distinct subdomains complete over the RAS solver
execution. For the synchronized RAS solver, all subdomains complete the same
number of local solves (even though this may not be needed as subdomains may
already have converged). In the asynchronous RAS, some subdomains complete a
much higher number of local solves, but the median of subdomain updates is smaller
than for the synchronized RAS solver. Even with a well-balanced problem and
machine setting, we see a difference of up to two in terms of the subdomain
update counts between the one-sided and two-sided communication paradigms. 

\section{Conclusion and Future work}
\label{sec:conc}

In this paper, we present a framework that allows to implement and compare
asynchronous and synchronous iterative methods for multi-node systems featuring
GPUs. The framework supports different communication schemes based on two-sided
MPI (synchronous) and one-sided MPI (asynchronous). In addition, both direct and
iterative local solvers are supported for a wide variety of modern HPC hardware,
including CPUs, NVIDIA and AMD GPUs. We implement both centralized and
decentralized convergence detection algorithms.

To study the benefits of using asynchronous iterative methods for sparse
problems, we realize an asynchronous version of the Restricted Additive Schwarz
algorithm and compare with the synchronous counterpart. The study is conducted
on the Summit HPC system with up to 16 nodes each equipped with 6 V100 NVIDIA
GPUs. We analyze the effect of different parameters such as partitioning,
overlap size, and convergence detection strategy for the synchronous and
asynchronous RAS solver. We also reveal that asynchronous RAS can for
optimal parameter choices complete up to 5.5$\times$ faster than its synchronous
counterpart.

As future work, we plan to implement and study more methods such as the
Optimized Restricted Additive Schwarz~(\cite{Magoules.Szyld.Venet.17}) that
overcomes the slow convergence of the RAS algorithm by communicating also the
boundary element derivatives. We also plan to investigate the use of iterative
local solves.

\bibliographystyle{unsrt}  
\bibliography{references}  

\end{document}